\documentclass[12pt,preprint,psfig]{aastex}

\shorttitle{The Halo of 4U 1624-490}
\shortauthors{Xiang, Lee, \& Nowak}

\def\1624{4U~1624-490} 
\def\chandra{{\it Chandra}}
\newcommand{\approxlt}{\mathrel{\mathop{\renewcommand{\arraystretch}{0}
\array{c} <\\\sim\endarray}}}
\newcommand{\approxgt}{\mathrel{\mathop{\renewcommand{\arraystretch}{0}
\array{c} >\\\sim\endarray}}}

\begin{document}

\title{Using the X-ray Dust Scattering Halo of 4U 1624-490 \\ 
to determine distance and dust distributions}

\author{Jingen Xiang\altaffilmark{1}, Julia C. Lee\altaffilmark{1}}
\affil{\altaffilmark{1}Harvard-Smithsonian Center for Astrophysics, 
60 Garden Street, Cambridge, MA, 02138}
\email{jxiang@cfa.harvard.edu; jclee@cfa.harvard.edu}
\and
\author{Michael A. Nowak\altaffilmark{2}}
\affil{\altaffilmark{2}Massachusetts Institute of Technology, Chandra X-ray Science Center 
\& Kavli Institute for Space Research, 77 Massachusetts Ave. NE80-6077, Cambridge, MA, 02139}
\email{mnowak@space.mit.edu}

\begin{abstract}
We present X-ray dust scattering halo results based on our 76~ks
\chandra\ ACIS-S/HETGS observation of the LMXB dipping source
4U~1624-490.  Through analysis of the halo light curves with 2-6~keV
spectra over the persistent and dipping periods, we estimate a
geometric distance of $\sim$15~kpc to \1624. We also fit halo radial
profiles with different ISM dust grain models to assess the location,
uniformity, and density of the halo. Our analysis shows that the dust
spatial distribution is not uniform along the line-of-sight; rather,
it is consistent with the spiral arm structure mapped in {\sc Hii}.
The large difference between the absorption Hydrogen column ($N_{\rm
H}^{abs} \sim 8 \times10^{22}\ \rm cm^{-2}$; probes all gas along the
line-of-sight) derived from broadband spectral fitting, and the
scattering Hydrogen column ($N_{\rm H}^{sca} \sim 4 \times10^{22}\ \rm
cm^{-2}$; probes only Galactic gas) derived from our studies of the
4U~1624-490 X-ray halo suggests that a large fraction of the column is
local to the X-ray binary.  We also present (and apply) a new method
for assessing the \chandra\ point spread function at large ($> 50''$)
angles, through use of the time delays from the observed dips.

\end{abstract}

\keywords{dust --- scattering --- X-rays: ISM --- sources: \1624}

\section{Introduction}
X-rays are not only absorbed but are also scattered by dust grains
when they travel through the interstellar medium (ISM).  This
scattering of X-rays from a source behind a dust containing cloud in
the ISM will lead to the formation of an X-ray scattering halo
surrounding the X-ray source. The properties of the halo depend upon
the size distribution and density of the dust grains, and on the
relative geometry of dust, X-ray source, and observer.

Such dust scattering halos were first discussed by
\citet{overbeck65},  but they were not observationally confirmed until
\citet{rolf83} observed the X-ray binary GX~339-4 with the {\it
Einstein} X-ray Observatory.  Since then, X-ray halos have been
studied via observations facilitated by X-ray satellites which include
{\it Einstein, ROSAT, BeppoSAX}, \chandra\ and {\it XMM-Newton}.  Thus
far, the most complete samples of  these studies have been presented
by \citet{predehl95} based on  {\it ROSAT} data, and by
\citet{xiang05} based on observations  with the \chandra\
ACIS-S/HETGS.  Of additional note, \citet{draine03} have analyzed
{\it ROSAT} observations of the halo  associated with the X-ray nova
V1974~Cygni~1992 to confirm that the interstellar dust model of
\citet[hereafter WD01]{wd01} is consistent with the observed X-ray
halos.  Furthermore, based on \chandra\ ACIS-I observations of  the
low-mass X-ray binary (LMXB) GX~13+1, \citet{smith02} reported that
both the grain model WD01 and the ``classical'' model of
\citet[hereafter MRN]{mathis77} can adequately reproduce its halo
radial profile.

X-ray halos are not limited to the aforementioned binaries.  Most
bright X-ray sources are surrounded by X-ray halos \citep{predehl95},
including one of the most extreme of the dipping sources, the {\it Big
Dipper\,} \1624, which is an atoll source \citep{lommen05} with an IR
counterpart of magnitude $K_{s}=18.3$ \citep{wachter05}.  The presence
of a halo in \1624 has been surmised via \textsl{BeppoSAX}
observations, which exhibit a soft excess above its several arcmin PSF
\citep{balucinska00}. The poor angular resolution of
\textsl{BeppoSAX}, however, prevents the direct extraction of the halo
radial profile at radii less than 100~arcsec, where the halo is much
more prominent.

In this paper, we present a focused study of the X-ray halo associated
with \1624, using the highest resolution (angular and energy) data to
date, as afforded by the \chandra\ ACIS-S/HETGS.  The $\sim$3~hours
long duration of the dips \citep{watson85,smale01, balucinska01} and
large obscuration of this compact object offer us the unique
opportunity to use the time delay of photons arriving from the halo to
determine the distance to \1624 (\S\ref{sec-theory} for theory and
\S\ref{subsec:dist} for data analysis), and compare with dust grain
models to assess composition and density along the line-of-sight (LOS,
\S\ref{subsec:halo}).  (The large obscuration may be due to the
accretion disk stream impacting the disk. A $\sim75\%$ obscuration has
been reported by \citealt{watson85} and \citealt{church95}; the
superior angular resolution of \chandra\, allows us to detect a 90\%
obscuration; \S\ref{subsec:dist}.)  We also present a new method for
assessing the \chandra\ Point Spread Function (PSF) at large angles
($\approxgt50''$), and compare the PSF measured by our technique with
{\sc Chandra Ray Tracing} ({\sc ChaRT}) predictions between 9--160$''$
(\S\ref{subsec-psf}).

\section{Theoretical and Historical Background} 
\label{sec-theory}

The theoretical calculations governing the observed halo surface
brightness and the time delay of a scattered photon with respect to an
unscattered one have been discussed extensively \citep[e.g.,][]{mauche84,
mathis91, trumper73}. Here, we briefly  describe the main points.

\begin{figure*}
\plottwo{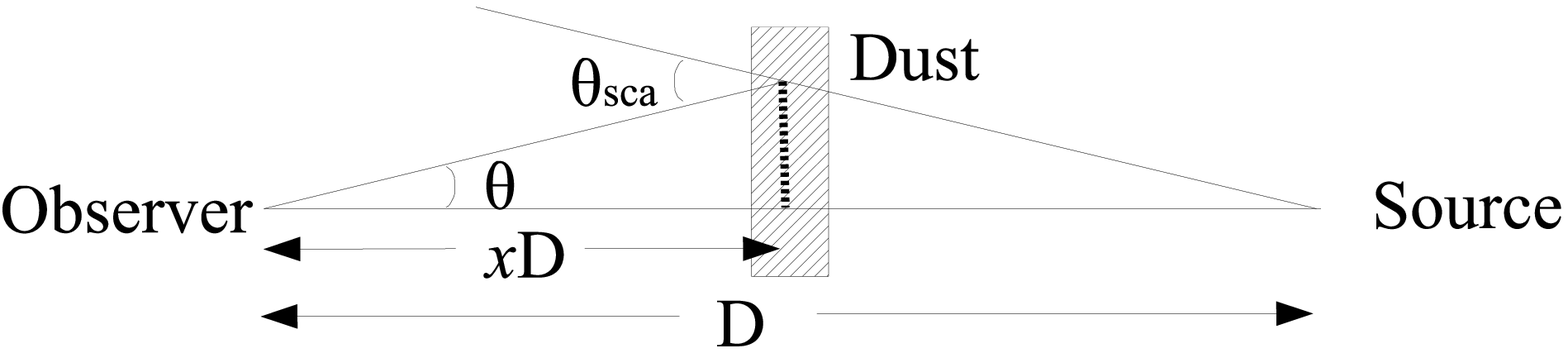}{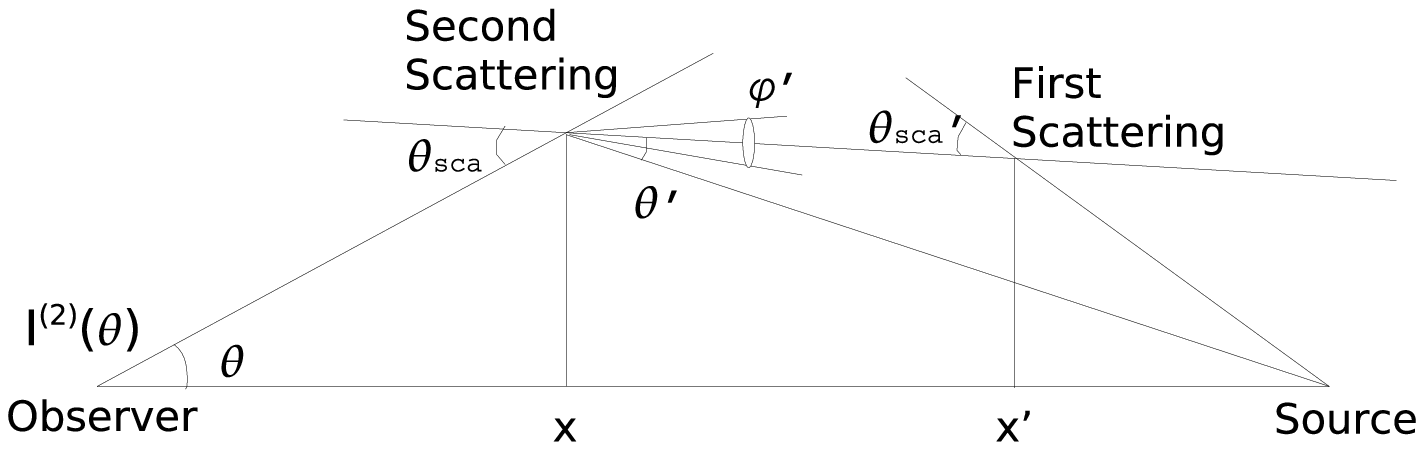}
\caption{Geometry of the X-ray-scattering process for  single 
(left) and double scattering (right).\label{geometry}}
\end{figure*}

As discussed by \citet{mauche84}, the differential cross section
in the Rayleigh-Gans approximation, coupled with the Gaussian approximation
for a spherical particle of radius $a$, can be described by
\begin{equation}
S(a,E,\theta_{sca})={{d\sigma_{sca}(a, E, \theta_{sca})} \over {d\Omega}} 
= 1.1 \times 10^{-12} \left({{2Z}\over {M}}\right)^{2}\left({{\rho}\over
{2}}\right) \left ( \frac{a}{\mu m} \right )^{6}\left[{{F(E)}\over 
{Z}}\right]^{2}\exp({-K^{2}\theta_{sca}^{2}}),
\end{equation} 
where $K=0.4575(E/keV)^2 (a/\mu m)^2$, and $\theta_{sca}$ is the angle of
scattering.  The mean atomic charge ($Z$),  molecular weight
($M$, in amu), mass density ($\rho$)  and atomic scattering factor
[$F(E)$,  from \citealt{henke81}] also factor into the calculation.
A photon of wavelength $\lambda$ will be scattered to a
typical angle
\begin{equation}
\theta=\lambda/\pi\ a\ \propto\ E^{-1}.
\end{equation}
For small angles (several arcmin), we can approximate the observed
angle by \textbf{$\theta$}$\approx(1-x) \theta_{sca}$, where $x=d/D$
is the relative distance defined to be the ratio of the distance from
scattering grain to observer ($d$) over the distance between source
and observer ($D$);  see Fig. \ref{geometry}.

The intensity of the observed first-order scattering halo,
$I_{sca}^{(1)}(\theta, E)$, depends on the X-ray flux $F_{X}(E)$ of
the point source (\1624 in our case), and is a function of both the
observed angle $\theta$ and the energy $E$.  The form of this
equation, as initially derived by \citet{mathis91}, is expressed as
\begin{equation}
\label{surface_I}
{I_{sca}^{(1)}(\theta, E) = F_{X}(E)}\
N_{\rm H}\int_{a_{min}}^{a_{max}}{da\
n(a)}{\int_{0}^{1}{dxf(x)(1-x)^{-2}}\times 
S\left(a,E,{\theta \over {1-x}}\right)},
\end{equation}
where $N_{\rm H}$ is the total hydrogen column density between the
observer and the X-ray source, $n(a)$ is the size distribution of the
dust grains, and $f(x)$ is the relative hydrogen density to the average
 total hydrogen density along the LOS at $xD$. For uniformly distributed dust, 
$f(x)\equiv 1$, and eq.~(\ref{surface_I}) can take on
the more explicit form
\begin{equation}
\label{surface_simple}
{I_{sca}^{(1)}(\theta, E) \propto F_{X}(E)}\
N_{\rm H}\int_{a_{min}}^{a_{max}}{da\ n(a)\ \left( \frac{a}{\mu m} \right )^{6}}\
{{{\rm erfc}(R)}\over {R}},
\end{equation}
where $R=K\theta_{sca}=K({\theta}/{1-x})$  and
${\rm erfc}(R)=({2}/{\sqrt{\pi}})\int_{R}^{\infty}dt\ \exp{(-t^{2})}$.

\citet{mathis91} further established that multiple scattering is
likely important if there is enough scattering optical depth to see an
appreciable halo. As discussed by those authors, as well as by
\citet{predehl96}, for a fixed energy $E$ doubly scattered radiation
at the position of the observer can be described via
\begin{eqnarray}
\label{surface_double}
I_{sca}^{(2)}(\theta, E)=F_{X}(E)\ N_{\rm H}\ \int_{0}^{1} dx\ f(x) \int_{x}^{1}
{{dx'\ f(x')} 
\over {(1-x')^2}} \int_{0}^{\infty}\theta'd\theta'\int_{0}^{2\pi}d\phi
\nonumber \\
\times\int_{a_{min}}^{a_{max}} da\ n(a)\ S(a,E,\theta_{sca})
\times\int_{a_{min}}^{a_{max}}da'\ n(a')\ {S(a^{'},E,\theta_{sca}^{'})},
\end{eqnarray}
where 
\begin{equation}
\theta_{sca} = \theta'^{2} + 2\theta'{2}\ sin\phi\ {\theta \over {1-x}} 
+ {{\theta^2} \over {(1-x)^2}},
\end{equation}
\begin{equation}
\theta_{sca}^{'2}=\theta'\ {{1-x}\over{1-x^{'}}}.
\end{equation}

For a typical scattering optical depth of $\tau_{sca}\approx0.5$, the
doubly scattered radiation dominates the multiple scattering terms at
several  arcmin such that higher order ($>2$) scatterings can  be
largely neglected.

Other factors, in particular the ISM dust composition and grain size
distribution, affect our overall  determination of the halo
properties.  Several different models exist for describing the
composition and size distribution of ISM dust grains. The two most
commonly used  are the grain model of \citet[WD01]{wd01}, and the
``classical'' one of  \citet[MRN]{mathis77}. The details of these two
models are different,  although they are both based on IR
observations. The MRN model assumes both graphite and silicate grains
with size distributions: $n(a)\propto a^{-3.5}$, for
$a_{min}<a<a_{max}$, where $a_{min}=0.005\ \mu m$ is the same for both
grain types, whilst $a_{max}=0.25\ \mu m$ for silicate grains and
$a_{max} \approx 0.25 - 1\ \mu m$ for graphite grains. In contrast,
the WD01 model, which includes very small carbonaceous grains
($a<0.005\ \mu m$) and larger grains  ($a>1\ \mu m$), is comparatively
more complex than the MRN model.  To illustrate, the size of the
carbonaceous grains extends to more than 1 $\mu m$, while the number of
grains decreases sharply with size -- see \citet{wd01} for details.

\section{Observation}
We observed \1624 on 2004 June 4 (MJD: 53160.26813, ObsID: 4559) with
the \chandra\ High Energy Transmission Grating Spectrometer (HETGS)
for 76 ks, covering one binary orbit. To reduce pileup, the
observation was performed using a reduced 1.7 sec frame-time and
one-half sub-array corresponding to 512 columns per CCD. Fig
\ref{fig-lc} shows that our observation encompasses $\sim$2.7 hr total
of dipping periods (3 dipping events with durations respectively of
about 3.5~ks, 2.3~ks and 4.0~ks).

\section{Data Analysis}

We used {\sc ciao}~3.3 with {\sc caldb}~3.2 to extract HETGS spectra
of the source for the persistent (non-dipping) and dipping periods
(Fig.~\ref{fig-lc}). The 2$-$6~keV halo light curve between 3$''$ and
20$''$ is extracted from the CCD~S3~0$^{th}$ order data, while the
point source light curve in the same energy band is extracted from the
dispersed data of the gratings (1$^{st}$ order data). We note that the
halo light curve, between 3$''$ and 20$''$, will have $\sim40\%$
contamination from the bright point source photons scattered into the
wings of the PSF.  {\sc ciao}~3.3 is also used to extract the surface
brightness of the 0$^{th}$ order data for the persistent and dipping
periods. (Note that in order to ensure that neither the halo nor the
PSF radial profiles are contaminated by the dispersed spectra of the
HETGS, throughout we have excluded pie slices of the regions from the
zeroth-order image that contain the MEG and HEG arms.)  Our analysis
technique for the results presented in \S\ref{subsec:dist} and
\S\ref{subsec:halo} are best illustrated by the flowcharts of Figs.
\ref{distance_chart}~and~\ref{halo_chart}, respectively.

\subsection{The broadband spectrum as determined from $1^{st}$ order HETGS data}
In order to determine the distance (\S\ref{subsec:dist}) and
distributed Hydrogen column  (\S\ref{subsec:halo}) along the LOS to
\1624, it is necessary that its light curve, and the total  flux
covering the entire $\sim$~76~ks spectrum, the dipping phase
($\sim$~12~ks), and persistent phases (persistent 1 + persistent 2,
$\sim$~59~ks) are estimated
over the same energy bands. (This corresponds to 2$-$6~keV over 200~eV
incremental steps for this paper; see \S\ref{subsec:dist} for
details.).  Furthermore, because there is a direct proportionality
between the source  spectrum and the X-ray halo intensity such that
any uncertainties associated with the spectral analysis will map onto
the halo analysis,  we take care to use spectra which suffer from the
least amount of pile-up in our continuum modeling.

\subsubsection{The time averaged continuum covering the binary cycle}  
\label{sec:fullcont}
The best fit model based on HEG$\pm 1$ fits to the time-averaged
$\sim 76$~ks spectrum between 1.5-10.0~keV is
$1.37_{-0.05}^{+0.08}$~keV  blackbody plus $\Gamma=1.0_{-0.9}^{+0.6}$
power-law modified by $N_{\rm H}=7.6_{-0.6}^{+0.9}\times 10^{22}$
cm$^{-2}$.  Not surprisingly, these parameters are similar to that
derived from the persistent period which is discussed  in more depth
subsequently.

\subsubsection{The persistent phase continuum}  
\label{sec:nondipcont} 
The maximal count rate for the HEG$\pm$1 spectra during this phase is
$7.8 \times 10^{-3}$ counts per pixel per frame time, corresponding to
a maximal pileup $\sim2\%$ at $\sim$5 keV.  The maximal count rate per
pixel in the MEG$\pm$1 is about two times that of HEG$\pm$1,
indicating a maximal pileup of $\sim4\%$.  Therefore, we have taken
care to use the nearly pileup free HEG$\pm$1 spectra to derive the
flux for the different bands used in our analysis of
\S\ref{subsec:dist} and \S\ref{subsec:halo}. The halo intensity is
proportional to $E^{-2}$ such that it decreases
sharply with increasing energy.  We therefore retain as many 
low energy ($\sim 2$~keV) halo photons as possible for our halo studies.
We further note that we derive broad band spectral parameters
based on the 1.5-10~keV spectral region, whilst only the 2-6~keV flux
(measured in 200~eV steps) is used in order to match our halo studies
of \S\ref{subsec:halo}.

\begin{figure*}
\epsscale{0.9}
\plottwo{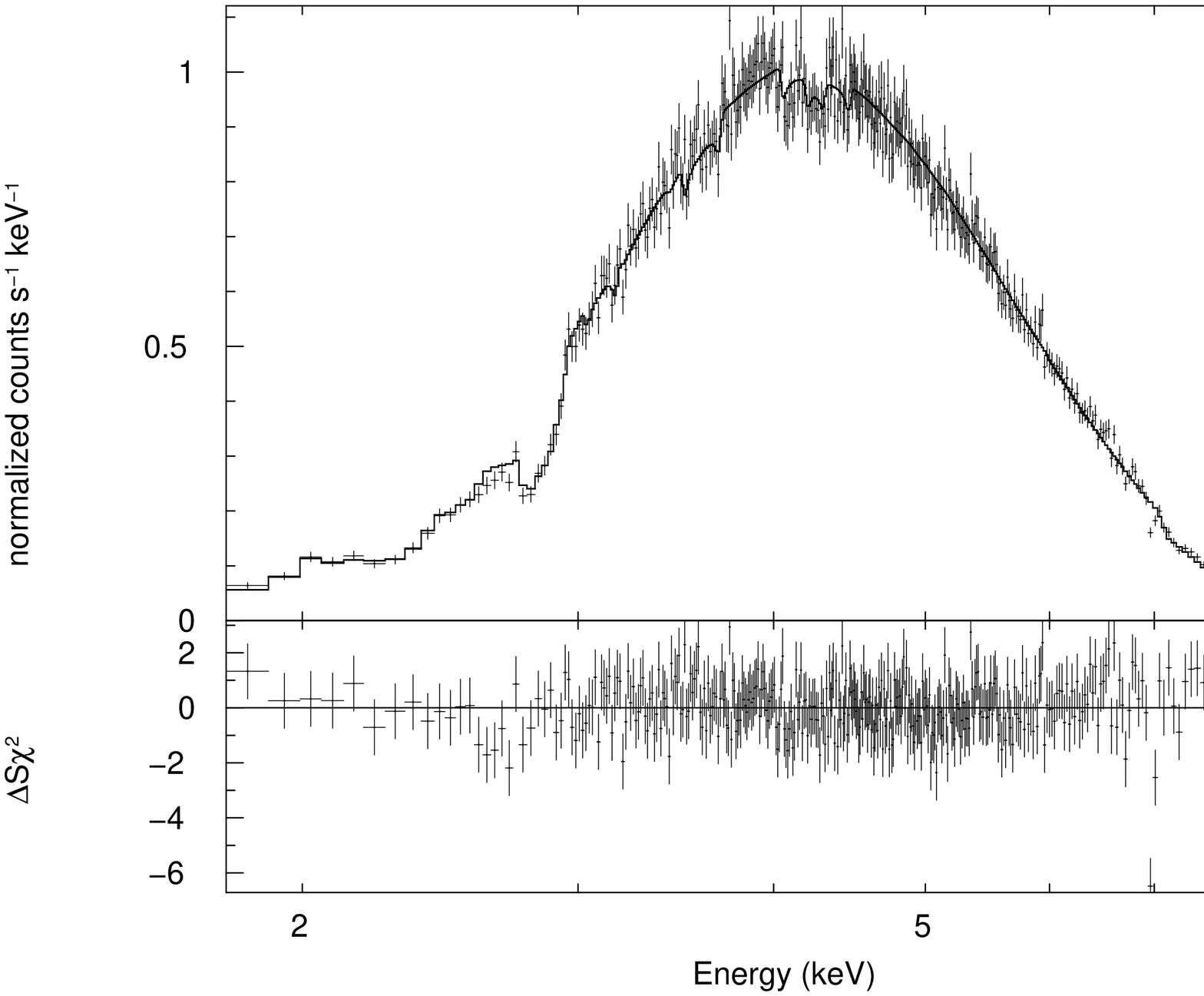}{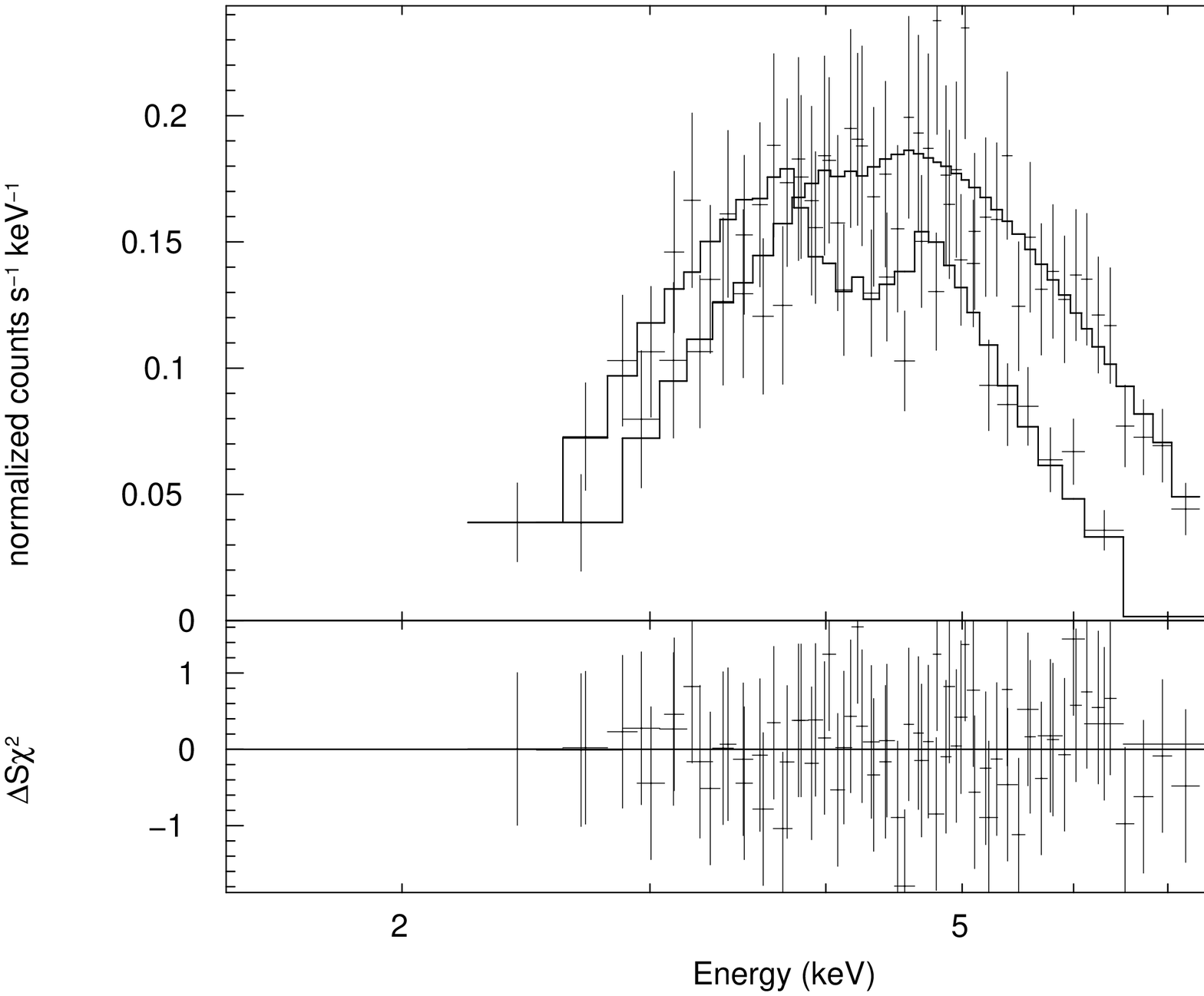}
\caption{The best fit models over-plotted on the broad band spectra of 
\1624 during the  persistent phase (HEG$\pm$1; left)
and dipping periods (HEG$\pm$1 and MEG$\pm$1; right).  \label{fig-spectra}}
\end{figure*}

\begin{figure*}
\plotone{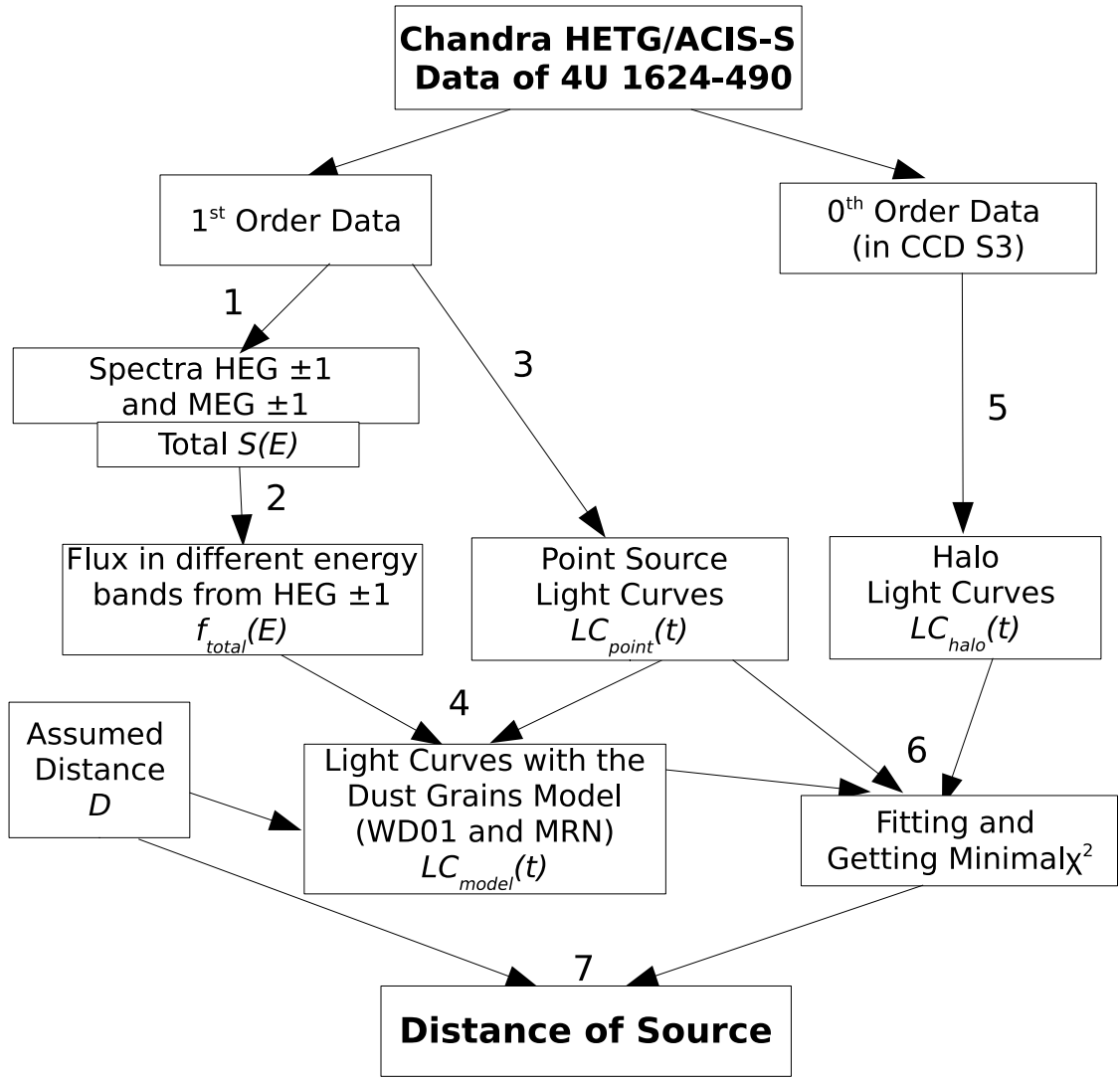}
\caption{Flowchart showing the data analysis process for determining the distance
to \1624. \label{distance_chart}}
\end{figure*}

Using
\footnote{http://heasarc.gsfc.nasa.gov/docs/xanadu/xspec/index.html}
{\sc xspec 12.2.1} \citep{arnaud96}, we fit the broadband source
continuum HEG$\pm1$ spectrum of the persistent phase with various
combinations of powerlaw, blackbody, diskbb and thermal bremsstrahlung
modified by the Tuebingen-Boulder ISM absorption model
\citep{wilms00}.  (Because of the high count rate of the source
spectrum, the data were binned to require at least 400 counts in each
energy bin.)  The fitting results are listed in Table
\ref{tab-spectra}. We find that both
Model~1:~absorption$\times$(blackbody+powerlaw) and
Model~2:~absorption$\times$(diskbb+powerlaw) fit the data well.
However, the latter gives an unreasonable photon index of $\Gamma =
12_{-4}^{+8}$, whilst the former agrees better with the parameters
reported by \citet{balucinska00} based on {\it BeppoSAX} data, by
\citet{smale01} based on {\it RXTE} data, and by \citet{parmar02}
based on {\it XMM-Newton} data, barring a flatter photon index in our
fitting. The steepness of $\Gamma$ appears strangely tied to choice of
cold absorption model (i.e. {\sc wabs} versus {\sc tbabs}), which we
will explore in more depth in our forthcoming paper on the high
resolution \chandra\ spectrum of \1624.  For present purposes, it
suffices that Model~1 describes the broadband continuum spectrum well
(see Fig.~\ref{fig-spectra}), such that we can confidently extract
flux values over the incremental 200~eV steps between 2-6~keV for our
calculations of \S\ref{subsec:halo}.

\subsubsection{The dip-phase continuum} 
\label{sec:dipcont}
We also fit the MEG$\pm$1 and HEG$\pm$1 spectrum of the dip phase with
many of the same models. Due to the lower overall count rate during
this phase, we were only able to rebin these data to require
S/N$\sim$8 bin$^{-1}$.  Both the MEG$\pm$1 and HEG$\pm$1 dip spectra
are pileup free.  We find the best fit for this broad band dip
continuum spectrum to be a $\Gamma = 0.7\pm0.5$ power-law modified by
$N_{\rm H}=9.05_{-0.19}^{+0.25}\times 10^{22}$ cm$^{-2}$ cold
absorption. Fluxes in 200~eV steps were also obtained in the same
2$-$6 keV energy band to compare with the persistent periods.

\begin{deluxetable}{llllc}
\tablecaption{Spectral fits to non-dip continuum.}
\tablehead{ \colhead{Model} &
\colhead{$N_{\rm H}$ ($10^{22}$~cm$^{-2}$)} &
\colhead{kT (keV)} & 
\colhead{$\Gamma$} & 
\colhead{$\chi^{2}$/dof}}
\startdata 
blackbody + Power law  & 7.8$_{-0.6}^{+0.8}$  & 1.34$_{-0.05}^{+0.09}$
& 1.1$^{+0.5}_{-1.0}$ & 364/379\\ 
Disk blackbody + Power law   & 9.04$^{+0.27}_{-0.19}$  & 2.57$\pm$0.06 & 12$^{+7}_{-4}$ & 356/379\\ 
Disk blackbody   & 8.76$\pm$0.13  & 2.64$\pm$0.05 &  &373/381\\ 
Bremsstrahlung & 10.02$\pm$0.14  & 10.3$_{-0.5}^{+0.6}$ & & 442/381\\ 
Blackbody   & 6.60$\pm$0.11  & 1.54$\pm$0.02 &  & 428/381\\ 
Power law   & 10.77$\pm$0.18 &  & 1.90$\pm$0.04 & 510/381\\ 
\enddata
\label{tab-spectra}
\tablecomments{These results come from fits to the \chandra\
ACIS-S/HETGS HEG$\pm$1. The MEG$\pm$1 is not used since it suffered
from more pileup than HEG$\pm$1. Errors are 90\% confidence. All
models listed are modified by the Tuebingen-Boulder ISM absorption
model of \citet{wilms00}, as noted in the $N_{\rm H}$ values.}
\end{deluxetable}

\subsection{Using the halo and point source light curves to determine distances} 
\label{subsec:dist}
The scattering photon travels longer distances than the unscattered
one. The delay time $dt$ is given by
\begin{equation}
\label{eq:delay_time}
dt = 1.15\ h\ (D/1\ kpc)\ (\theta/1\ arcmin)^2{{x}\over{1-x}},
\end{equation}

Thus measurement of a time delay between the point source and the halo
yields direct information about the distance to the dust cloud and the
source. The $\sim$~3 hr long dipping of \1624 provides an excellent
opportunity to search for such a delay. \citet{trumper73} and
\citet{xu86} proposed methods for using this behavior to measure the
distance to variable X-ray sources, and \citet{predehl00}  presented a
successful geometric distance determination to Cyg X-3  based on
\chandra\ ACIS-S/HETGS observations.

The time delay $dt$ is only dependent on the location of scattering,
but the halo surface brightness is determined by the size, position,
and composition of the dust grains, as well as the source flux and
scattering hydrogen column.  Therefore, the spectrum of the source,
the dust grains model and the spatial distribution of the grains is
implicitly factored into eq.~(\ref{eq:delay_time}).

The halo intensity is proportional to $E^{-2}$ such that it decreases
dramatically at the very high energies.  Fig.~\ref{fig-spectra} shows
that the majority of the \1624 counts are contained within the 2-6~keV
energy band. Therefore, we restrict our comparative analysis of the
halo and point source light curve exclusively to this band, where
fluxes are estimated over 200~eV steps based on the continuum model
described in \S\ref{sec:fullcont}.

In order to avoid pileup effects, the bright point source light curve
is extracted directly from the dispersed $1^{st}$ order data of the
HETGS, where the count rate is a factor of $\sim 4$ lower than that of
the non-dispersed $0^{th}$~order data.  In contrast, we extract the
(lower flux) halo light curve between 3$''$ and 20$''$ from the
$0^{th}$ order data, where the minimum angle is restricted to 3$''$ to
mitigate pileup.  We estimate that while pileup at 3$''$ is less than
2\%, between 3$''$--20$''$ it is $\ll$~1\% so that we can safely
neglect it for our analysis.  The maximum angle restriction of 20$''$
was chosen to reduce multiple scattering effects, which if not
factored in properly can lead to over-predictions of the time delay.
Alternatively, we could have restricted the energy band to 3--6\,keV,
for example.  However, this not only would have reduced the degree of
multiple scatterings, it would have greatly decreased the halo
intensity, which scales as $E^{-2}$.  Thus we opted for the angular
restriction with an energy that covers well the observable point
source continuum.  We estimate that the total halo intensity between
3$''$--20$''$ associated with the second-order scattering is less than
5\% that of the first-order scattering in the 2-6~keV energy band of
interest.  While we account for 2$^{nd}$ order scattering effects in
our analysis, we wanted to make sure that it would not play a major
role in affecting our final results.

\begin{figure*}
\epsscale{0.9}
\plotone{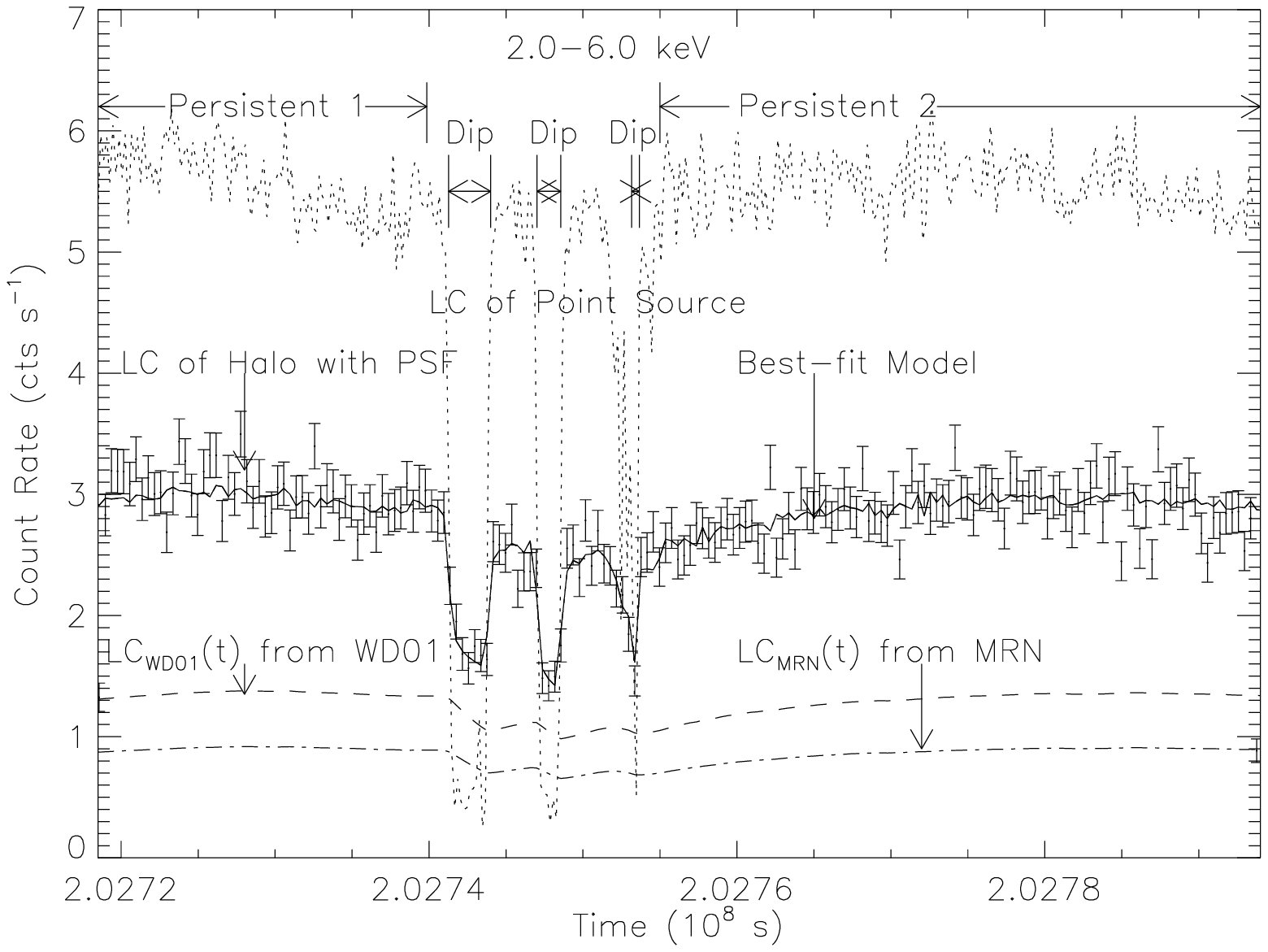}
\caption{The 2--6\,keV halo and point source light curves. The
top-most dotted line is the point source light curve for \1624. The
combined halo and PSF light curve between 3$''$ and 20$''$ is shown,
with error bars, in the middle. The overplotted solid line is the
``best fit'' ($\chi^{2}$ statistics) for the halo light curve based on
$LC_{WD01}(t)$ (dashed line) convolved with the PSF lightcurve.  To
facilitate clearer viewing, we multiply the observed halo plus PSF
light curve by a factor 4, the $LC_{WD01}(t)$ from model WD01 by a
factor of 3, and the $LC_{MRN}(t)$ from model MRN by a factor of 2.
\label{fig-lc}}
\end{figure*}

A comparison of the halo and point source light-curve
(Fig.~\ref{fig-lc}) reveals a mean delay of $\sim$1.6~ks for source
photons arriving from the halo that is at 3$''$--20$''$ from \1624. It
is interesting to note that the dips we detect show as much as
$\approxgt$~90\% blockage of the compact object (Fig.~\ref{fig-lc}),
compared, for example, to the 75\% reported previously in {\it EXOSAT}
\citep{watson85}, {\it Ginga} \citep{jones89}, {\it BeppoSAX}
\citep{balucinska00} and {\it RXTE} \citep{smale01} observations.
This is attributed to the fact that the superior imaging capabilities
of \chandra\ can better separate the source light-curve from
contaminated light from the halo. Using the time delay, and
eq.~(\ref{eq:delay_time}), a simple estimate for the distance to \1624
can be derived. If we assume a mean value for the fractional
distance:~$\bar x={1\over 2}$ and a mean effective angle
$\bar\theta=10.3''$ derived using
\begin{equation}
\int_{\theta_{min}}^{\theta_{max}}{\theta d \theta}
\exp\left[ -Ka^2E^2 \left ( {\theta\over{1-\bar x}} \right )^2 \right ]
=\bar\theta(\theta_{max}-\theta_{min})
\exp \left [ {-Ka^2E^2 \left ({\bar\theta\over{1-\bar x}} \right )^2} \right ]
\end{equation}
where $\theta_{min}=3$ arcsec and $\theta_{max}$=20 arcsec,  we obtain
a distance to \1624 to be about 13.1 kpc.  In the subsequent
discussion, we discuss a more rigorous approach  for deriving the
distance.

Following the prescription presented by \citet{predehl00}, we
initially generate model light curves $LC_{\rm WD01}(t)$ (using the
WD01 dust model) and $LC_{\rm MRN}(t)$ (MRN dust model) for the halo
(Fig.~\ref{fig-lc}), according to eqs.~(\ref{surface_I}) \&
(\ref{eq:delay_time}), assuming a uniform spatial distribution of dust,
for an initial guess of the distance $D$=10~kpc.  [$LC_{\rm WD01}(t)$
and $LC_{\rm MRN}(t)$ will hereafter be referred to generically as
$LC_{model}(t)$.] Then we use $LC_{model}(t)$ together with the light
curve of the point source $LC_{point}(t)$ to fit the light curve of
the halo $LC_{halo}(t)$, allowing for the ratio,
$LC_{model}(t)$:$LC_{point}(t)$ to vary until a minimal $\chi^{2}$ is
found.  This is iterated for input distances ranging between
8$-$20~kpc in 0.1~kpc steps. The best fit distance to \1624 derived
from the WD01 and MRN models for uniformly distributed dust are
respectively $D_{\rm WD01}=15.0_{-2.6}^{+2.9}$~kpc and $D_{\rm
MRN}=10.2^{+2.4}_{-1.4}$~kpc.  (Errors are quoted at 90\% confidence
based on variations in $\chi^2$ with $D$.)  Within errors, both models
appear to give consistent values for the distance.  However, it may be
of interest to note that the distance derived from the MRN model is
somewhat sensitive to the allowed maximum size $a_{\rm max}$ for the
dust grains. Since the maximum size for carbonaceous grains in the MRN
model is not specified, we tried to test our distance determination by
changing the maximum size for carbonaceous grains in the MRN
model. For $a_{\rm max}=0.25\ \mu$m, we obtain $D_{\rm
MRN}=10.2^{+2.4}_{-1.4}$~kpc, as noted above, in comparison to $D_{\rm
MRN}=15.0^{+3.0}_{-2.6}$~kpc for $a_{\rm max}=0.42\ \mu$m grain sizes.
The latter distance estimate is comparable to the value obtained based
on the WD01 model.

The WD01 model likely provides a more realistic present day picture of
the dust grains in the ISM.  \citet{smith02}, however, reported that
the MRN model is better than the WD01 model in fitting the halo radial
profile of GX~13+1, based on observations using the \chandra\,
ACIS-I. R.~Smith (priv. comm., Dec. 2006) notes, however, that this
may be attributable to systematic errors. In a recent reanalysis of
GX~13+1, based on a new \chandra\, calibration, Smith finds that the
WD01 model may be preferred (although the differences between the WD01
and MRN models are still small).
Therefore, for our distance determinations, we are still inclined to
accept the value of $D_{\rm 4U1624}=15.0_{-2.6}^{+2.9}$~kpc based on
fits using the WD01 model.  This is consistent with the 10-20 kpc
value derived by \citet{christian97}, based on a method which compared
the hydrogen column densities from their spectral fitting to an
exponential distribution model of hydrogen in the Galaxy.


Furthermore, since dust grains between us and the point source are
{\it unlikely} to be uniformly distributed, we check whether and to
what extent our distance determinations are affected by unevenly
distributed dust. Accordingly, in our modeling, we vary the dust
placement in additional fits, e.g., x=0.5-1, 0.4-1, etc., where x=0
and x=1 correspond respectively to our position and that of the source
(see Fig.~\ref{fig-spiralarm}). We find our distance estimates to be
robust to these changes, {\it as long as} the dust is uniformly
distributed {\it near} the source. This is because the distance
estimates are based primarily on small angle scatterings, where the
observed halo is attributed primarily to dust near the source (see,
e.g., Fig~4 of \citealt{mathis91}).  This finding is consistent with
our results based on fits to the halo profile for determining column
densities in \S\ref{subsec:halo}.

\subsection{Using the halo radial profile to determine the $N_{\rm H}$ spatial distribution} 
\label{subsec:halo}
Fits to the halo light curve show that $\sim$40\% of the photons  come
from the point source as far out as 20$''$.
While \chandra\ has very good imaging resolution, the halo brightness
is not much greater than the PSF (see Fig. \ref{psf_halo},
right). Therefore, a good understanding of the telescope PSF is
necessary for optimal results.

\subsubsection{The {\it Chandra} PSF and halo of \1624}
\label{subsec-psf}
As discussed by \citet{smith02}, the {\sc ciao} tool {\sc mkpsf} and
the \chandra\ raytrace model {\sc SAOsac} recreate the observed core
of the PSF well but underestimate the wings of the PSF. The same is
true of the tool {\sc ChaRT} since it also uses {\sc SAOsac}.  Column
(d) of Table \ref{tab-psf} lists the comparison between the PSF
($PSF_{ChaRT}$) from the {\sc ChaRT} simulation and the one
($PSF_{HerX1}$) derived from the \chandra\ ACIS-S/HETGS observation of
the almost halo-free source, Her X-1 (ObsID 6149).  Parameters derived
from the Her X-1 observation are used in its {\sc ChaRT}
simulation. As such, we can confidently use {\sc ChaRT} to assess the
PSF behavior at small ($\approxlt50''$) angles. For angles larger than
50$''$, we subsequently present a new method for assessing the
\chandra\ PSF by using the observed dips in \1624.

\begin{figure*}
\plotone{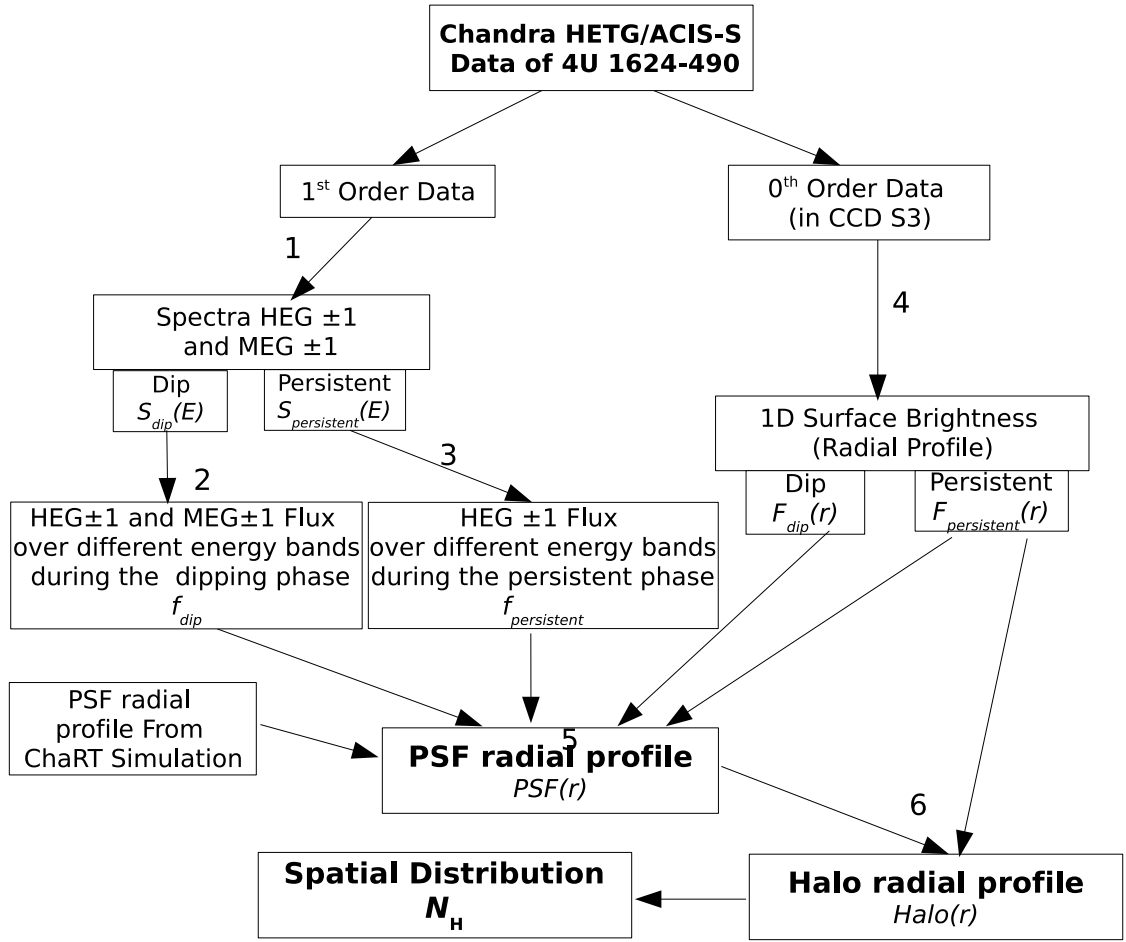}
\caption{Flowchart showing the data analysis process for determining
the PSF radial profile and halo radial profile which are used to
assess distributions of LOS $N_{\rm H}$. \label{halo_chart}}
\end{figure*}

\begin{figure}
\epsscale{0.7}
\plotone{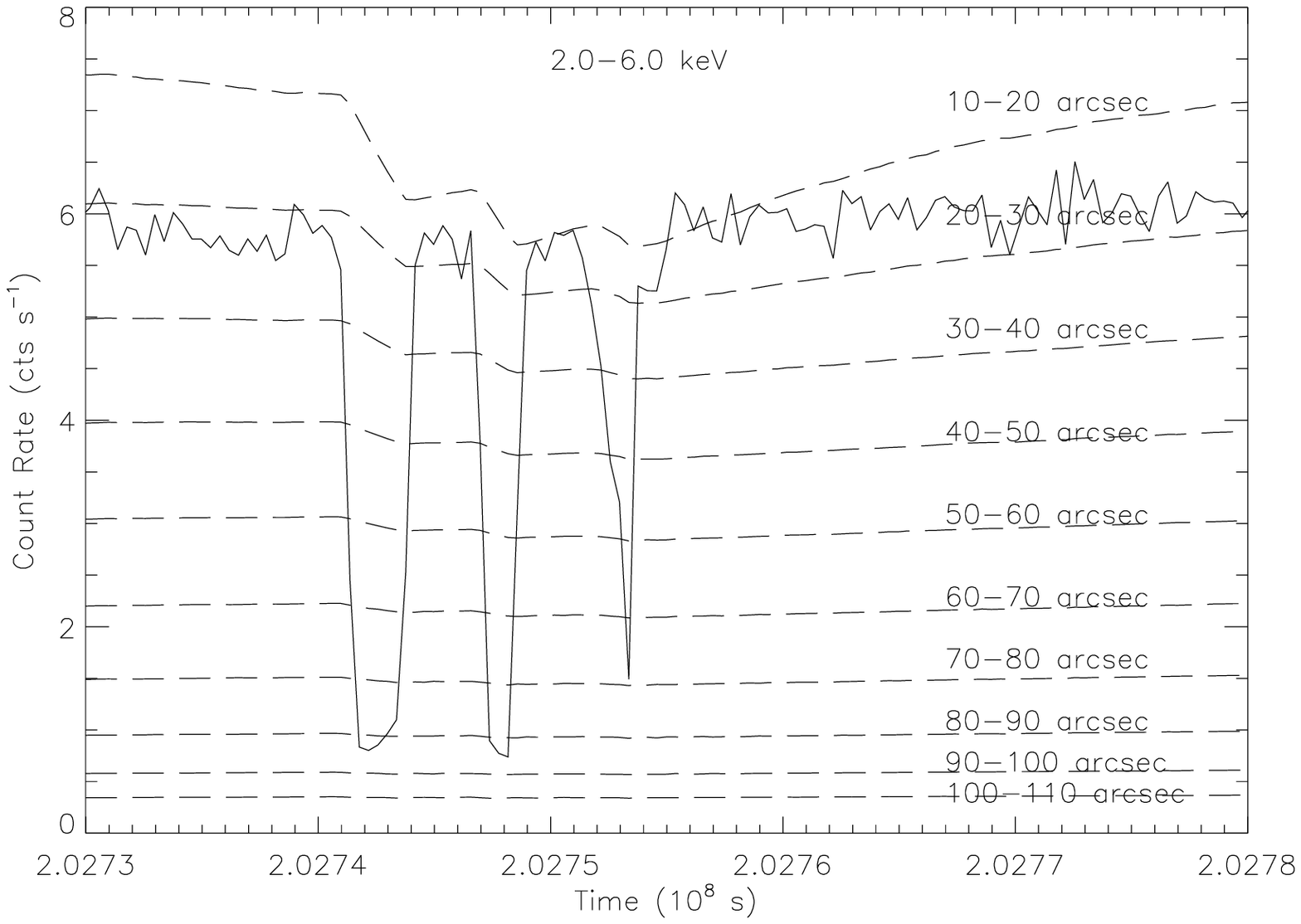}
\caption{The theoretical halo light curves derived from the WD01 model (dashed)
at different angular distances based on the observed point source (\1624) light
curve (solid). The halo light curves are arbitrarily multiplied by 
some constant factor for easier viewing. \label{fig-lc_model}}.
\end{figure}

Fig. \ref{fig-lc_model} shows the WD01 halo model light curve over
different angular distances compared against the point source light
curve.  It is clear that at increasingly larger angles,  the flux
behavior of the point source is reflected less in the halo light curve
(Table~\ref{tab-psf}).  This is in part due to a ``smearing" effect at
large angles as well as the increased delay between the 
photons arriving from the point source and those arriving from the halo.
Capitalizing on this observed effect of the dipping phenomenon on the
halo light curve, we discuss a prescription for assessing the
\chandra\  PSF for \1624 at large angles. First we define:

\begin{equation}
\label{eq-psf}
PSF(r) = [I_{persistent}(r)-I_{dip}(r)] {F_{persistent} \over (F_{persistent} -
F_{dip})},
\end{equation}
where $I_{dip}(r)$ and $I_{persistent}(r)$ are respectively, the
surface brightness over different radii during the time periods when
\1624 is dipping and when it is not.  Similarly, $F_{dip}$ and
$F_{persistent}$ refer to the point source flux during these
periods\footnote{As the halo dips affect the halo surface brightness,
especially shortly after the dips, throughout we calculate
$I_{persistent}$ solely from persistent phase 1, i.e., the $\approx
20$\,ksec before the dips.  However, for the point source flux,
$F_{persistent}$, we use both persistent phases 1 and 2.}.  Based on
this, we can then define the surface brightness of the halo~:
\begin{equation}
\label{eq-halo}
I_{\rm halo}(r) = I_{persistent}(r) - PSF(r) = \left [ I_{dip}(r) F_{persistent} 
- I_{persistent}(r) F_{dip} \right ] / (F_{persistent} - F_{dip}).
\end{equation}

To track properly the spectra of the dip and persistent periods and
determine how these spectra are reflected in the behavior of the
energy dependent PSF, we use eq.~(\ref{eq-psf}) to calculate the PSF
over the energy range 2--6\,keV in 200\,eV steps.  Note that this
equation implicitly assumes no time-dependence of the halo flux.  This
is, of course, inaccurate, especially at small angular distances from
the source. It becomes more accurate, however, at large angular
distances where the dipping behavior is ``smeared out'' in the halo
lightcurves (see Fig.~\ref{fig-lc_model}).  Thus, we expect
eq.~(\ref{eq-psf}) to be less dominated by systematic uncertainties at
larger angular distances.  Large angular distances are also where {\sc
ChaRT} estimates of the PSF become more problematic.

To assess the systematic errors in our determination of the PSF radial
profile, we use eqs.~(\ref{surface_I}) and (\ref{eq:delay_time}) to
calculate the theoretical halo light curves [LC$_{\rm mod}$(t)] at
different ranges of angular distances, as shown in
Fig.~\ref{fig-lc_model}.  Specifically, as described in
\S\ref{subsec:dist}, we used the lightcurve of the persistent source
with different presumed models for the source distance and dust
distribution, and chose a model that minimizes the $\chi^2$ for the
fit to the halo lightcurves. The halo intensity averaged over the dip
phases [i.e., essentially the estimate embodied in
eq.~(\ref{eq-halo})] can than be compared to the halo intensity in the
persistent phases (i.e., the ``true'' steady state halo flux).  This
estimate is presented in column (a) of Table \ref{tab-psf}.  As
expected, for radii near the point source, selecting the periods
during the dips slightly underestimates the persistent halo flux, and
hence leads to a systematic overestimation at small angular radii of
the PSF as calculated via eq.~(\ref{eq-psf}) [column (c) of
Table~\ref{tab-psf}].

As shown in Table~\ref{tab-psf}, for radii $<50''$, {\sc ChaRT} yields
less than 13\% uncertainties for the PSF, while for radii $>50''$,
using eq.~(\ref{eq-psf}) yields uncertainties $<10\%$ for the PSF.
The relative uncertainty in the halo flux is approximately given by
the fractional uncertainty in the PSF multiplied by the ratio of the
PSF flux to the halo flux [i.e., columns (c) multiplied by
column (b) or columns (d) multiplied by
column (e) from Table~\ref{tab-psf}].  Thus, by using the {\sc ChaRT}
PSF estimate for radii $<50''$, and eq.~(\ref{eq-psf}) for radii
$>50''$, the systematic uncertainty in the halo profile should be $<
3\%$ everywhere.

\begin{center}
\begin{deluxetable}{cccccc}
\tablecaption{Estimation of the Systematic Error of PSF and Halo. \label{error}} 
\tablehead{ 
\colhead{radius ($''$)} &
\colhead{${{I_{halo}^{wd01, dip}(r)- I_{halo}^{wd01}(r)} \over I_{halo}^{wd01}(r)}^{a}$} &
\colhead{${PSF_{data}(r) \over I_{halo}^{data}(r)}^{b}$} & 
\colhead{${{\Delta PSF_{data}(r)} \over PSF_{data}(r)}^{c}$} & 
\colhead{${{PSF_{HerX1}(r) - PSF_{ChaRT}(r)} \over {PSF_{HerX1}(r)}}^{d}$} &
\colhead{${PSF_{ChaRT}(r) \over I_{halo}^{ChaRT}(r)}^{e}$}}
\startdata
09-20   & -0.12  & 0.77 & 0.18 & 0.02 & 0.36\\
20-30   & -0.07  & 0.51 & 0.16 & 0.08 & 0.23\\
30-40   & -0.05  & 0.46 & 0.13 & 0.13 & 0.20\\
40-50   & -0.04  & 0.39 & 0.11 & 0.13 & 0.19\\
50-60   & -0.03  & 0.32 & 0.09 & 0.24 & 0.16\\
60-70   & -0.02  & 0.24 & 0.09 & 0.30 & 0.13\\
70-80   & -0.01  & 0.36 & 0.04 & 0.32 & 0.11\\
80-90   & -0.01  & 0.25 & 0.04 & 0.38 & 0.11\\
90-100  & -0.003  & 0.16 & 0.02 & 0.46 & $<$0.10\\
100-160 & $<-0.003$  & $\sim$0.30 & $<$0.01 & $>0.40$ & $<$0.10\\
\enddata 
\tablenotetext{a}{$I_{halo}^{wd01}(r)$ and $I_{halo}^{wd01, dip}(r)$
mean the halo intensity for the persistent and (averaged) dip
phases, respectively, derived from fitting the energy-dependent
lightcurves with the WD01 model.}
\tablenotetext{b}{$PSF_{data}(r)$ is the PSF intensity derived from
eq.~(\ref{eq-psf}), while $I_{halo}^{data}(r)$ is the halo intensity
derived from eq.~(\ref{eq-halo}).}
\tablenotetext{c}{${\Delta PSF_{data}(r)}$ is an estimate of
systematic error in the (persistent phase) PSF due to the
underestimate of the derived halo intensity [i.e., column
(a)]. Specifically, we use the estimate 
$\Delta PSF_{data}(r)/PSF_{data}(r)$ $\approx$ 
$[[I_{halo}^{wd01}(r)-I_{halo}^{wd01,dip}(r)]/I_{halo}^{wd01}(r)]$ 
$[F_{persistent}/(F_{persistent} - F_{dip})]$
$[I_{halo}^{data}(r)/PSF_{data}(r)]$.}
\tablenotetext{d}{$(PSF_{HerX1} - PSF_{ChaRT})/PSF_{Her X1}$
means the systematic error of PSF intensity from
{\sc ChaRT} simulations compared with the Her X-1 observational data.}
\tablenotetext{e}{$I_{halo}^{ChaRT}(r)=I_{persistent}(r)-PSF_{ChaRT}(r)$ is the halo intensity with the {\sc ChaRT} simualtion PSF.}
\tablecomments{{\it The data in this table is
is used solely to discern the relative accuracy of our PSF estimates, and are not used
in the calculation of either the PSF or halo radial profiles.}}
\label{tab-psf}
\end{deluxetable}
\end{center}

\begin{figure*}
\epsscale{1.0}
\plottwo{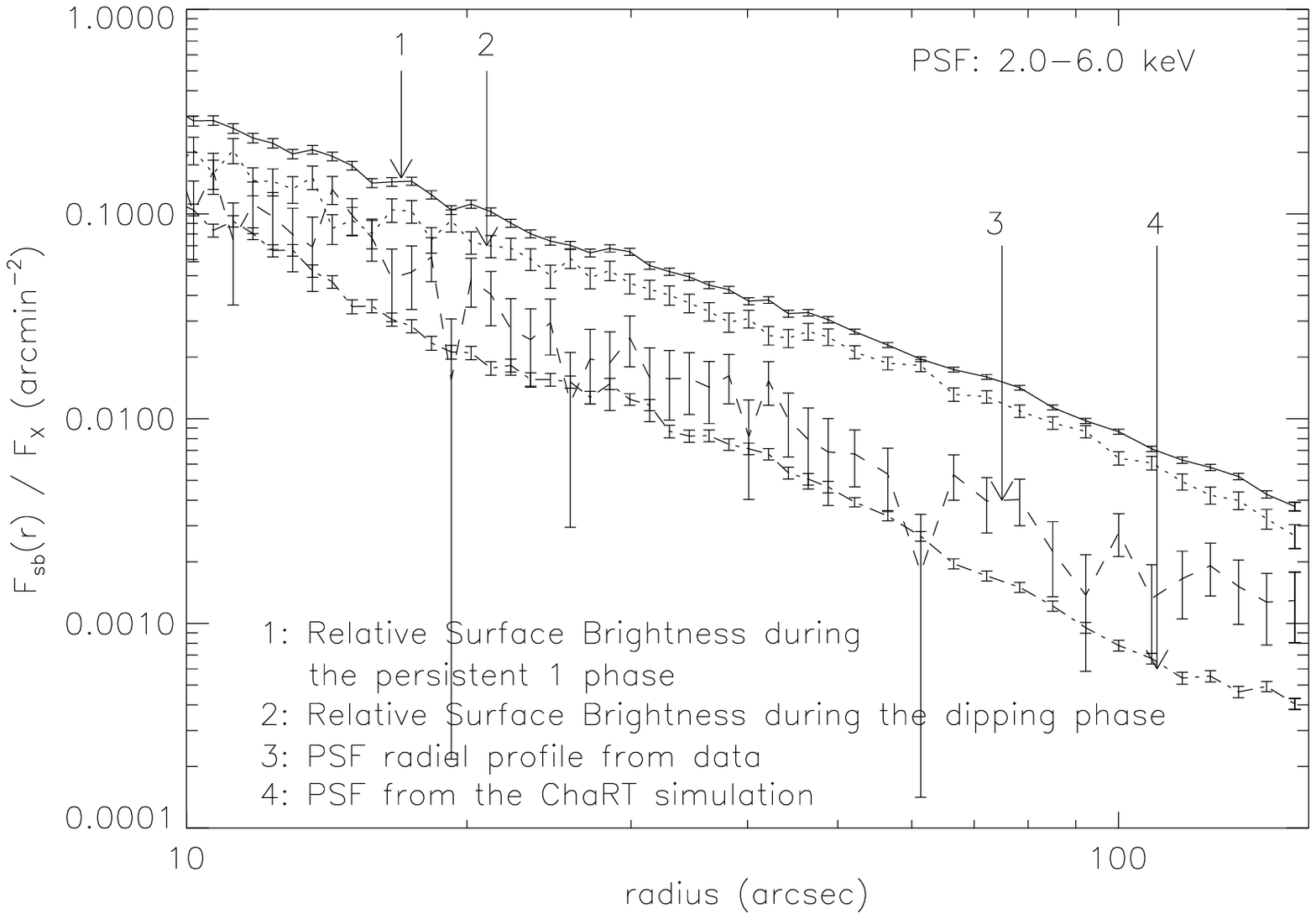}{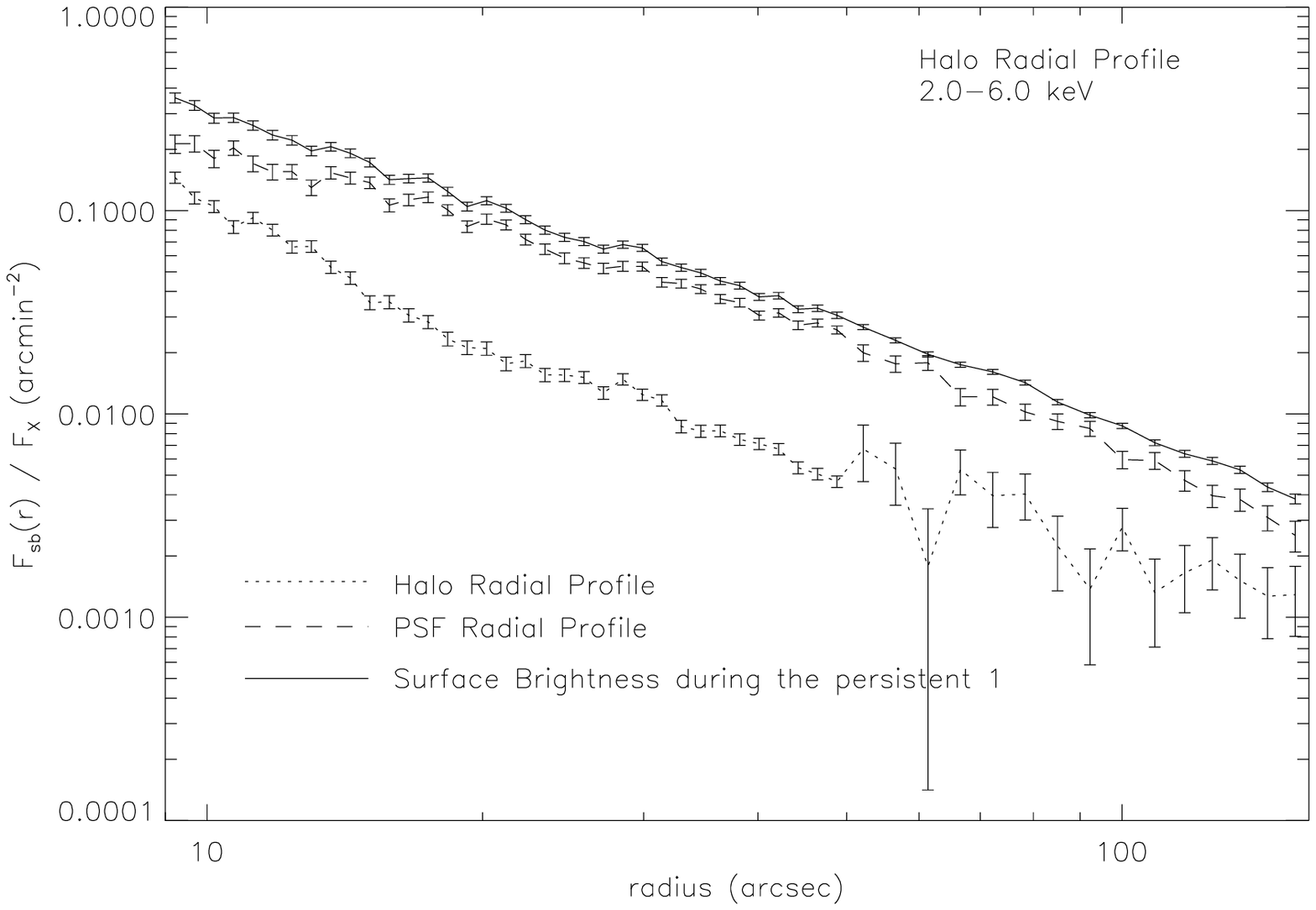}
\caption{PSF (left) and halo (right) radial profile of 4U
1624-490.\label{psf_halo}}
\end{figure*}

\subsubsection{Fits to the halo radial profile}
\label{subsec:halo_fit}
Having gained some understanding of the PSF behavior, we use the WD01
and MRN  models to fit the halo radial profile using the fitting codes
developed by  \citet{smith02}, as also applied to the \chandra\ halo
studies presented in  \citet{xiang05}. See also Fig.~\ref{halo_chart}
for a flow chart of the analysis  process. We confine our halo fitting
to angular distributions corresponding  to radii 9$''$ to 160$''$
where the extreme of pileup effect is $<0.5\%$ at $\approxgt9''$ and
decreases with angular distance. Therefore, we can assume a pileup
free halo for our analysis.

\begin{figure}
\epsscale{0.5} \plotone{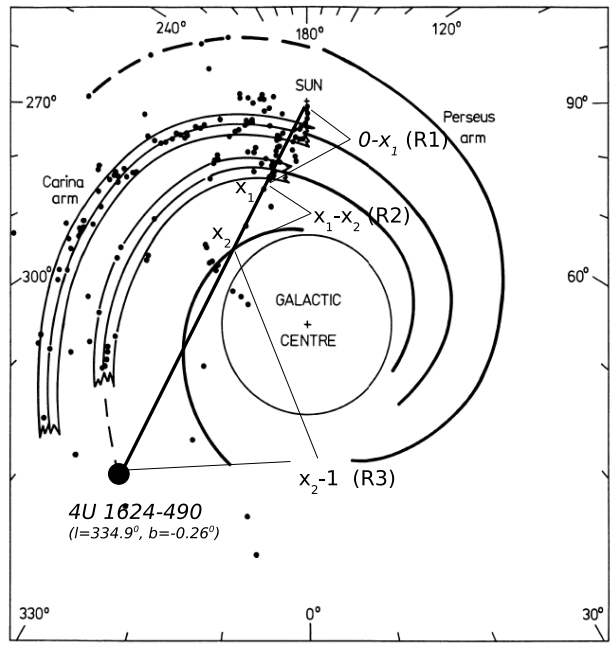}
\caption{The spiral arm structure of the Milky Way Galaxy and the
position of \1624 in it. The figure is recreated from
\citet{caswell87}. The positioning and the source is based on the
distance derived in \S\ref{subsec:dist} \label{fig-spiralarm}}
\end{figure}

In order to assess dust properties and distribution, for our initial
fitting, we assume uniformly distributed dust between x=0 (us) and x=1
(\1624; see Fig. \ref{fig-spiralarm}, as recreated from
\citealt{caswell87} and annotated to show the distance of \1624 from
us).  Based on this scenario, we find that neither the MRN model (with
$a_{max}(graphite)=0.42\ \mu$m; $\chi^{2}/dof=141/49$ for a best fit
$N_{\rm{H}}=4.44_{-0.08}^{+0.08}\times10^{22}\ \rm{cm}^{-2}$) nor the
WD01 model ($\chi^{2}/dof=171/49$ with
$N_{\rm{H}}=3.40_{-0.06}^{+0.06}\times10^{22}\ \rm{cm}^{-2}$) provides
good fits to the halo profile.  Both models under-predict the halo
surface brightness at $>$80$''$ by $\sim$50\%. We also fit the halo
radial profile to include the second-order scattering and find a 50\%
substantial improvement in our fits.  (We remind the reader that the
angular regions probed in this study are much larger than the distance
determination study of \S\ref{subsec:dist}, where multiple scattering
effects were found to be negligible.)  While the fits improved
somewhat, they are still statistically unacceptable with
$\chi^{2}/dof\sim91/49$ for the WD01 model, and
$\chi^{2}/dof\sim89/49$ for the MRN model of
$a_{max}(graphite)=0.42\,\mu$m.

\begin{figure}
\epsscale{0.5} \plotone{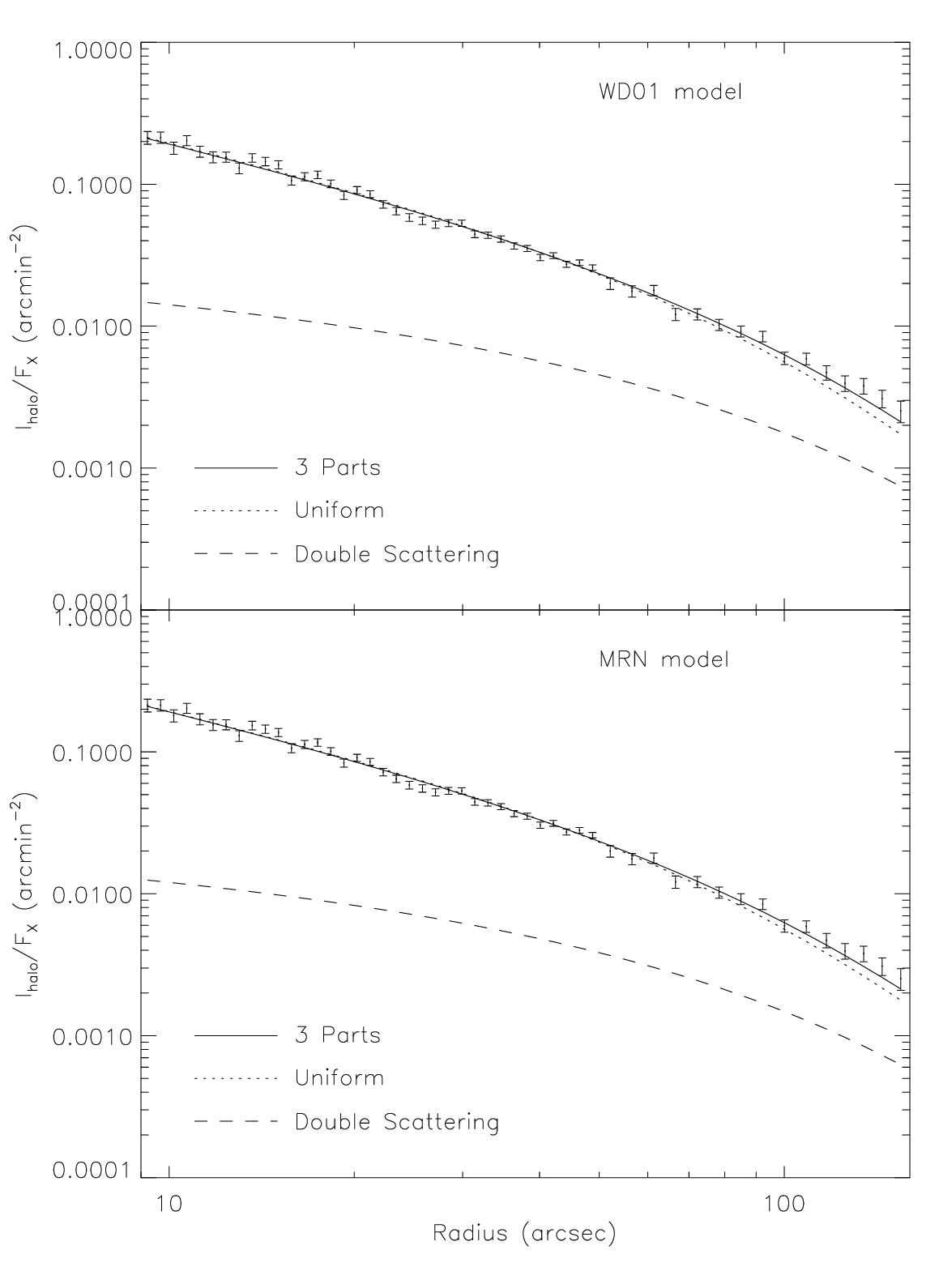}
\caption{X-ray halos observed at 2.0-6.0~keV, fit with dust grains model
WD01 (top) and MRN (down). The solid line is the model where the dust
spatial distribution is divided into three parts, which is marked
in the Fig. \ref{fig-spiralarm}, and the dotted line is the model with
uniformly distributed dust, respectively. The dashed line is the double scattering 
intensity model with 3 parts dusts. \label{halo_fit}}
\end{figure}

\begin{deluxetable}{cccccc}
\tabletypesize{\scriptsize} 
\tablecaption{Halo fits for deriving $N_{\rm H}$ based on a three parts dust distribution scenario.} 
\tablehead{ 
\colhead{Model} &
\colhead{$N_{\rm{H}}^{R1}(10^{22}$ cm$^{-2}$)} &
\colhead{$N_{\rm{H}}^{R2}(10^{22}$ cm$^{-2}$)} & 
\colhead{$N_{\rm{H}}^{R3}(10^{22}$ cm$^{-2}$)} & 
\colhead{$N_{\rm{H}}^{tot}(10^{22}$ cm$^{-2}$)} &
\colhead{$\chi^{2}/d.o.f$}   } 
\startdata 
WD01 & $1.58\pm0.15$ & $0.00^{+0.18}_{-0.00}$ & $2.04_{-0.09}^{+0.08}$ & $3.62_{-0.06}^{+0.06}$ & 65/47 \\
MRN & $2.08^{+0.19}_{-0.20}$ & $0.00^{+0.23}_{-0.00}$ & $2.73\pm0.11$ & $4.81^{+0.08}_{-0.08}$ & 65/47\\
\enddata 
\label{halo_NH}
\end{deluxetable}

Having determined that the uniformly distributed dust between us and
the point source \1624 is an unlikely scenario, we investigate whether
the models are sensitive to patchy distributions.  Accordingly, based
on Fig.~\ref{fig-spiralarm}, we roughly divide the dust distribution
along our LOS into three parts~:~x~=~0.0-0.20 (Region 1: R1),
0.20-0.40 (R2), 0.40-1.0 (R3), corresponding to a distance $d$
relative to us of,~respectively, 0$-$3.0~kpc (R1), 3.0$-$6.0~kpc (R2),
and 6.0$-$15.0~kpc (R3), as seen in Fig. \ref{halo_fit}.  While dust
is assumed to be smoothly distributed in each of these regions, the
quantity is allowed to vary independently.  Based on this, fits to the
halo radial profile using the WD01 model yield Hydrogen column
densities for each region (Table~\ref{halo_NH}).  Compared to fits
assuming uniformly distributed dust between x=0$-$1, $\Delta \chi^2 =
26$ for 47 $d.o.f.$ ($\chi^{2}_{\nu}=1.38$), i.e., $>$99$\%$
confidence for 2 additional parameters, according to the F-test.
Similarly, fits based on three parts with the MRN model give $\Delta
\chi^2 = 24$ for 47 $d.o.f.$ ($\chi^{2}_{\nu}=1.38$), i.e., again
$>$99\% confidence for 2 additional parameters. The second-order
scattering is accounted for in these fits by adding the numerically
integrated value of the second-order scattering to the first-order
scattering value during fitting of the halo radial profile. The
scattering hydrogen column density $N_{\rm H_{MRN}}=4.8\times 10^{22}\
{\rm cm}^{-2}$ derived from the MRN model is consistent with the
result ($N_{\rm H}=5.0\times 10^{22}\ {\rm cm}^{-2}$ corresponding to
$\tau=2.4$) obtained by \citet{balucinska00}.  We also notice,
however, that a smaller $\tau=1.8$ obtained by \citet{trigo06} is more
consistent with the result $N_{\rm H_{WD01}}=3.6\times 10^{22}\ {\rm
cm}^{-2}$ derived from the WD01 model.

The derived hydrogen column density  $N_{\rm H_{WD01}}^{\rm
R1}=(1.58\pm0.15)\times 10^{22}$~cm$^{-2}$ (translated into a particle
density  $n_{\rm R1} \sim$1.7 cm$^{-3}$) for the two nearby spiral
arms encompassed within R1, is consistent with the result derived from
the X-ray halo of  Circinus~X1 by \cite{xiang05} over similar
distances ($N^{\rm Cir X-1}_{\rm H_{WD01}}=1.4\times10^{22}$
~cm$^{-2}$ at $\sim$~3~kpc from us) along a similar LOS ($l=322^{0}$,
$b=0.04^{0}$). It should be noted that these two values do not  need
to be exactly equal since the LOS to \1624 and the one to Circinus~X-1
pass through different regions of the same two spiral arms.  For R2
(the region between spiral arms 2 and 3), we estimate an upper limit
to the particle density $n_{\rm R2} \sim$0.2 cm$^{-3}$) based on a
90\% confidence value for $N_{\rm H}^{\rm R2} \le 1.8 \times
10^{21}$~cm$^{-2}$ (Table~\ref{halo_NH}). Combined, the hydrogen
particle density in R1 is $n_{\rm R1}\sim$1.7 cm$^{-3}$,  twice as
high as the average density along the entire  LOS to \1624. The best
halo radial profile fits are shown in  Fig. \ref{halo_fit}.


While the absorption hydrogen column density includes both the
Galactic  ISM and the gas nearby the source, the scattering hydrogen
column density only comes from the Galactic interstellar
medium. Therefore, our finding that the scattering $N_{\rm H}$
derived  independently from  the WD01 and MRN models is much less than
the absorption $N_{\rm H}$  derived from spectral fitting, implies
that there is significant absorption intrinsic to  the source. If we
naively take the difference between the best fit  value $N_{\rm
H}^{abs} \sim 8 \times 10^{22}$ cm$^{-2}$ of the broadband
HETGS spectra due to Galactic and source ISM  and $N^{sca}_{\rm
H} \sim 4 \times10^{22}$ cm$^{-2}$  derived from our halo studies, a
$N_{\rm H}\approxlt 4 \times10^{22}$~cm$^{-2}$  can conceivably be
local to the source, possibly put there by the  stellar wind of the
companion.

\section{Summary}

\begin{itemize}
\item We improve upon previous distance estimates for \1624 by making use
of the delay time of the halo photons relative to the bright point
source photons to obtain  $D_{\rm
4U1624}=15.0_{-2.6}^{+2.9}$~kpc. This is consistent within the errors
of the 10-20~kpc estimates by \citet{christian97} using a different
technique.

\item We find that varying dust distributions {\it will not} affect
our distance determination to \1624 except possibly for a scenario
where there is no dust within $\sim$7.5~kpc of the source
(x=0.5-1.0). This extreme scenario does not match the $N_{\rm H}$
derived from our halo fitting results so it can be ignored as
potentially problematic for our distance estimates.

\item Using the extreme dipping behavior of \1624, we discuss a new
method for  estimating the \chandra\ PSF at large angles ($>50''$). In
a comparison  with {\sc ChaRT} estimates for the PSF, we find that if
we estimate the PSF using {\sc ChaRT} at $<50''$, and our method at
$>50''$, we can limit the errors  associated with our halo analysis
to  $\approxlt$ 3\% over the angular 9$''-$160$''$ region.

\item Varying dust distribution {\it does} affect the derived column
densities. A simple estimate based on our halo fits imply the hydrogen
particle density in the spiral arms is $n_{\rm R1}\sim$1.7 cm$^{-3}$,  and
the one between two spiral arms $n_{\rm R2}<$0.2 cm$^{-3}$.

\item For the future, larger field-of-view and high throughput observations
combined with on-going \chandra\, studies will allow us to better
diagnose the scattering of dust near and far from us to reveal more
detailed spatial information.

\end{itemize}

\noindent Note : After submission of this paper, \cite{iaria06} posted
a preprint presenting the spectral analysis of our HETGS data for
\1624 which is currently public.  Our complete analysis of these data
will be presented in a forthcoming publication.

\acknowledgments{We wish to thank Randall Smith for insightful
suggestions. This work  was funded by the NASA / \chandra\ grant
GO4-3056X -- we are thankful for its support. We are also grateful to 
the Harvard University Clark fund for research support.}

\end{document}